# Fundamentals of Molecular Integrals Evaluation

Justin T. Fermann and Edward F. Valeev

# Contents

# 1 Introduction

At the dawn of quantum mechanics physicists realized that not every partial differential equation nature has to offer can be solved analytically. Even the most fundamental equation of quantum mechanics, the Schrödinger equation, can be solved analytically only for very few systems. As Dirac put it in 1929, "the underlying physical laws necessary for the mathematical theory of a large part of physics and the whole of chemistry are thus completely known, and the difficulty is only that the exact application of these laws leads to equations much too complicated to be soluble". Thus, to apply quantum mechanical concepts to generic molecular systems one has to find some other way of solving the complicated equations.

One approach is to solve differential equation numerically, via introduction of a finite grid. This reduces the problem to having to solve a (large) system of linear equations, a problem well suited for solving on digital computers. However, high accuracy via this approach can be achieved only with non–uniform grids, with high density of points around the nuclei.

An alternative approach, more intuitively obvious for chemists familiar with the concept of atomic orbitals (physicists familiar with Bloch functions), is to expand the solution in terms of functions that behave like the solution, atomic orbitals in the chemist's case (plane waves in the physicist's). Again, this simplifies the problem down to solving matrix equations, with one complication being that the matrix elements become more or less non–trivial to compute. The matrix elements of the Hamiltonian operator consist of several types of *molecular integrals*, and the evaluation of these integrals is the main focus of this document.

Of course, one has to anticipate what the exact solution looks like, but it turns out that the fundamental properties of atomic and molecular wavefunctions, such as nuclear and electron–electron cusps, exponential decay at infinity, etc. are known and it is relatively easy to find trial functions to satisfy them. Atom-centered basis sets help to describe the critical region around the nuclei efficiently, hence these have been used in a majority of quantum chemical studies of molecules. (Homework problem – find a study that didn't utilize atom-centered functions...)

The choice of a particular functional form for the expansion is more difficult. If one had to solve the Schrödinger equation for the hydrogen atom in finite basis, the natural choice would be to choose functions of the form

$$\phi = r^L e^{-\zeta r} Y_L^M(\theta, \phi) \tag{1.1}$$

that resemble the well–known discrete–spectrum solutions found in any textbook on quantum mechanics. Introduced by Slater, this type of basis functions is particularly well suited for atomic calculations. However, the applicability Slater–type functions to other than trivial molecular problems is hindered by the enormous computational complexity of the resulting expressions for matrix elements of the Hamiltonian.

Boys found a more tractable choice in 1950[1] as he introduced Gaussian–type functions of the form

$$\phi = x^l y^m z^n e^{-\alpha r^2} \tag{1.2}$$

and they have been with us ever since. Much simpler expressions for the matrix elements more than compensate for improper behavior of Gaussians at the origin and infinity. Indeed, an *s*-type Gaussian (a Gaussian functions with $l+m+n$ equal 0) is smooth at the origin, whereas an *s*-type Slater-type function has a cusp at the origin (non-zero derivative with respect to $r$). Also, Gaussian-type functions decay with $r$ *much* faster that Slater-type functions. However, a fact that a given Slater-type functions can be well represented as a linear combination of only few Gaussians with different exponents was noticed early on, thus STO–$n$G basis sets were introduced, in which a single Slater-type function is represented as a linear combination of $n$ Gaussians.

One important note to make before we move on to mathematical details is the value of efficient algorithms for computing integrals. Normally, in a typical high accuracy calculation only a small portion of CPU time is spent in computing molecular integrals, and the major part is spent in computing wavefunction parameters. However, the situation has changed dramatically since the late 1980s. Direct methods which do not store the integrals but recompute them as needed put the integrals evaluation process into a spotlight again. OK, enough said, on to the integrals!

## 2 Elementary Basis Function Analysis

### 2.1 Normalization

As we have mentioned, the standard basis functions used in *ab initio* theory are Gaussian functions, or linear combinations thereof. There are two types of Gaussians: Cartesian and spherical harmonic. The functional expression for the unnormalized primitive Cartesian Gaussian–type functions is worth rewriting here:
$$\phi_\mu(\mathbf{r}) = x^l y^m z^n e^{-\alpha r^2} \tag{2.1}$$
This is a basis function of angular momentum $(l+m+n)$,# centered at the origin, with orbital exponent $\alpha$. The term "primitive" denotes a fact that it is a single function. A linear combination of primitive Gaussians located at the same center
$$\phi_\mu(\mathbf{r}) = \sum_{i=0}^{c} x^{l_i} y^{m_i} z^{n_i} e^{-\alpha_i r^2} \tag{2.2}$$
is called a contracted Cartesian Gaussian function. Standard notation. A corresponding primitive spherical harmonic, or pure angular momentum, Gaussian is simply
$$\phi_\mu(\mathbf{r}) = r^L e^{-\alpha r^2} Y_L^M(\theta, \phi) \tag{2.3}$$
Note that we use capital letters for the angular momentum number and the projection of angular momentum on $z$-axis to distinguish those from

---
#Angular momentum is a slippery term here, some authors just use "orbital quantum number"

the exponents of $x$ and $y$ in Eqn (2.1). Hence, $L$ is related to $l + m + n$, but not identical. The rule is that any spherical harmonic Gaussian of angular momentum $L$ can be expressed solely in terms Cartesian Gaussians of angular momentum $L$. However, the reverse is not generally true. Similarities end there though. Just keep in mind that unless stated otherwise, terms "Gaussian" and "Gaussian function" will refer to Cartesian Gaussian–type functions throughout this document.

These atomic orbital–like basis functions need not be orthogonal to one another, but for later convenience, it would be nice to have them normalized. Thus impose the condition

$$\int \phi_\mu^*(\mathbf{r})\phi_\mu(\mathbf{r})d\mathbf{r} = 1. \tag{2.4}$$

Let's evaluate this integral. Assume a normalization constant of $N$ for $\phi_\mu$, and call (2.4) a self–overlap integral, SO.

$$\text{SO} = \int N^2 (x^l y^m z^n e^{-\alpha \mathbf{r}^2})(x^l y^m z^n e^{-\alpha \mathbf{r}^2}) d\mathbf{r} \tag{2.5}$$

$$= N^2 \int x^{2l} y^{2m} z^{2n} e^{-2\alpha \mathbf{r}^2} d\mathbf{r} \tag{2.6}$$

This integral over all space is separable when done in Cartesian coordinates.[#]
Using $r^2 = x^2 + y^2 + z^2$ and $d\mathbf{r} = dxdydz$, we get

$$\text{SO} = N^2 \int dx x^{2l} e^{-2\alpha x^2} \int dy y^{2m} e^{-2\alpha y^2} \int dz z^{2n} e^{-2\alpha z^2} \tag{2.7}$$

$$= N^2 I_x I_y I_z \tag{2.8}$$

Full derivation of these integrals ($I_x$, etc) is left to the reader (otherwise this would not be a learning experience). The result is

$$I_x = \int_{-\infty}^{\infty} dx x^{2l} e^{-2\alpha x^2} = \frac{(2l-1)!!\sqrt{\pi}}{(4\alpha)^l \sqrt{2\alpha}}. \tag{2.9}$$

---

[#]Spherical harmonic Gaussians aren't as nice in this respect, that's why normally all integrals are computed in the Cartesian basis and then transformed into the spherical harmonic basis.

Recall that $(2l-1)!! = 1 \cdot 3 \cdot 5 \cdots (2l-1)$. Thus

$$\text{SO} = N^2 \left[ \frac{(2l-1)!!(2m-1)!!(2n-1)!!\pi^{3/2}}{(4\alpha)^{(l+m+n)}(2\alpha)^{3/2}} \right] = 1 \qquad (2.10)$$

Rearranging that to solve for $N$, the normalization constant,

$$N = \left[ \left(\frac{2}{\pi}\right)^{3/4} \frac{2^{(l+m+n)}\alpha^{(2l+2m+2n+3)/4}}{[(2l-1)!!(2m-1)!!(2n-1)!!]^{1/2}} \right] \qquad (2.11)$$

This result is completely general – for uncontracted functions. As you might have guessed, computing normalization constants for contracted Gaussians is not much more difficult.

## 2.2 Products of Contracted Cartesian Gaussians

Examine some contracted $s$ functions. Let

$$\phi(\mathbf{r}) = N \sum_i^n a_i e^{-\alpha_i r^2} \qquad (2.12)$$

where $n$ is the number of primitive functions in the contracted function $\phi(\mathbf{r})$, and $a_i$ are the contraction coefficients. The product $\phi^*(\mathbf{r})\phi(\mathbf{r})$ can be written

$$\phi^*(\mathbf{r})\phi(\mathbf{r}) = N^2 \left[ \sum_i^n a_i e^{-\alpha_i r^2} \sum_j^n a_j e^{-\alpha_j r^2} \right]. \qquad (2.13)$$

Since the bracketed term contains a product of two polynomials, only two types of terms can result; the square of each uncontracted function and the biproducts of different uncontracted functions. Take an example where $n$ is three:

$$\begin{aligned}[a_1 e^{-\alpha_1 r^2} + a_2 e^{-\alpha_2 r^2} + a_3 e^{-\alpha_3 r^2}]^2 &= [a_1^2 e^{-2\alpha_1 r^2} + a_2^2 e^{-2\alpha_2 r^2} + a_3^2 e^{-2\alpha_3 r^2} \\ &\quad + 2\alpha_1\alpha_2 e^{-(\alpha_1+\alpha_2)r^2} + 2\alpha_1\alpha_3 e^{-(\alpha_1+\alpha_3)r^2} \\ &\quad + 2\alpha_2\alpha_3 e^{-(\alpha_2+\alpha_3)r^2}] \qquad (2.14)\end{aligned}$$

In this case, as in all others, there are only two types of terms of which the integral needs to be taken. They may be written and evaluated as

$$1. \int a_i^2 e^{-2\alpha_i r^2} d\mathbf{r} = a_i^2 \left(\frac{\pi}{2\alpha_i}\right)^{3/2} \tag{2.15}$$

$$2. \int 2a_i a_j e^{-(\alpha_i+\alpha_j)r^2} d\mathbf{r} = 2a_i a_j \left(\frac{\pi}{\alpha_i+\alpha_j}\right)^{3/2}. \tag{2.16}$$

It is realized that 1. above can be obtained by setting $i = j$ in 2., and henceforth only the general case needs to be considered. Generalizing to arbitrary $n$ is straightforward, and so the normalization of contracted Gaussian functions can proceed as

$$\begin{aligned}
\int \phi^*(\mathbf{r})\phi(\mathbf{r})d\mathbf{r} &= N^2 \pi^{3/2} \left[\frac{a_1^2}{(2\alpha_1)^{3/2}} + \cdots + \frac{2\alpha_1\alpha_2}{(\alpha_1+\alpha_2)^{3/2}} + \cdots \right] \\
&= N^2 \pi^{3/2} \sum_i^n \sum_j^n \frac{a_i a_j}{(\alpha_i+\alpha_j)^{3/2}} = 1,
\end{aligned} \tag{2.17}$$

thus the normalization constant for the entire contraction will be

$$N = \pi^{-3/4} \left[\sum_{i,j}^n \frac{a_i a_j}{(\alpha_i+\alpha_j)^{3/2}}\right]^{-1/2} \tag{2.18}$$

Contractions of Gaussians of arbitrary angular momentum are a bit worse, but if we assume all of the contracted functions to be of the same angular momentum

$$\phi(\mathbf{r}) = N\left[a_1 x^l y^m z^n e^{-\alpha_1 r^2} + a_2 x^l y^m z^n e^{-\alpha_2 r^2} + \cdots\right]$$

$$= N x^l y^m z^n \sum_i^n a_i e^{-\alpha_i r^2} \qquad (2.19)$$

$$\int \phi^*(\mathbf{r})\phi(\mathbf{r})d\mathbf{r} = N^2 \int x^{2l} y^{2m} z^{2n} \left[\sum_i^n a_i e^{-\alpha_i r^2} \cdot \sum_j^n a_j e^{-\alpha_j r^2}\right] d\mathbf{r} \quad (2.20)$$

$$\int \phi^*(\mathbf{r})\phi(\mathbf{r})d\mathbf{r} = N^2 \sum_{i=0}^n \sum_{j=0}^n a_i a_j \int \left[x^{2l} y^{2m} z^{2n} e^{-\alpha_i r^2} e^{-\alpha_j r^2}\right] d\mathbf{r} \quad (2.21)$$

The product in brackets in Eqn (2.21) we've encountered before. Analogous to Eqn (2.9), the general form for the integral in the double sum is

$$\int x^{2l} y^{2m} z^{2n} a_i a_j e^{-(\alpha_i + \alpha_j) r^2} d\mathbf{r} = a_i a_j \pi^{3/2} \frac{(2l-1)!!(2m-1)!!(2n-1)!!}{2^{(l+m+n)}(\alpha_i + \alpha_j)^{(l+m+n+3/2)}}. \quad (2.22)$$

The self overlap is then

$$\int \phi^*(\mathbf{r})\phi(\mathbf{r})d\mathbf{r} = \frac{N^2 \pi^{3/2}(2l-1)!!(2m-1)!!(2n-1)!!}{2^{l+m+n}} \sum_{i,j}^n \frac{a_i a_j}{(\alpha_i + \alpha_j)^{l+m+n+3/2}}. \quad (2.23)$$

Calling $l + m + n = L$, the angular momentum of the shell, and solving for $N$,

$$\int \phi^*\phi = \frac{N^2 \pi^{3/2}(2l-1)!!(2m-1)!!(2n-1)!!}{2^L} \sum_{i,j}^n \frac{a_i a_j}{(\alpha_i + \alpha_j)^{L+3/2}} = 1 \quad (2.24)$$

$$N = \left[\frac{\pi^{3/2}(2l-1)!!(2m-1)!!(2n-1)!!}{2^L} \sum_{i,j}^n \frac{a_i a_j}{(\alpha_i + \alpha_j)^{L+3/2}}\right]^{-1/2} \quad (2.25)$$

Before we approach one-electron integrals, we need to consider one very important result.

## 2.3 The Gaussian Product Theorem

The Gaussian Product Theorem states that the product of two arbitrary angular momentum Gaussian functions on centers **A** and **B** can be written as

$$\begin{aligned}
G_1 G_2 &= G_1(\mathbf{r}, \alpha_1, \mathbf{A}, l_1, m_1, n_1) G_2(\mathbf{r}, \alpha_2, \mathbf{B}, l_2, m_2, n_2) \\
&= \exp[-\alpha_1 \alpha_2 (\overline{\mathbf{AB}})^2/\gamma] \times \\
&\quad \left[ \sum_{i=0}^{l_1+l_2} f_i(l_1, l_2, \overline{\mathbf{PA}}_x, \overline{\mathbf{PB}}_x) x_P^i e^{-\gamma x_P^2} \right] \times \\
&\quad \left[ \sum_{j=0}^{m_1+m_2} f_j(m_1, m_2, \overline{\mathbf{PA}}_y, \overline{\mathbf{PB}}_y) y_P^j e^{-\gamma y_P^2} \right] \times \\
&\quad \left[ \sum_{k=0}^{n_1+n_2} f_k(n_1, n_2, \overline{\mathbf{PA}}_z, \overline{\mathbf{PB}}_z) z_P^k e^{-\gamma z_P^2} \right].
\end{aligned} \quad (2.26)$$

To show this, we first define the multiplicands as

$$G_1 = G_1(\mathbf{r}, \alpha_1, \mathbf{A}, l_1, m_1, n_1) = x_A^{l_1} y_A^{m_1} z_A^{n_1} e^{-\alpha_1 r_A^2} \quad (2.27)$$

$$G_2 = G_2(\mathbf{r}, \alpha_2, \mathbf{B}, l_2, m_2, n_2) = x_B^{l_2} y_B^{m_2} z_B^{n_2} e^{-\alpha_2 r_B^2}. \quad (2.28)$$

Here, $\mathbf{r}_A = \mathbf{r} - \mathbf{A}$, etc. For primary analysis, take the angular momentum of $G_1$ and $G_2$ to be zero, so

$$G_1 = e^{-\alpha_1 r_A^2}; \qquad G_2 = e^{-\alpha_2 r_B^2}. \quad (2.29)$$

These are unnormalized, but normalization can be calculated as in Section 2.1. It would be convenient if this product could be written as a third Gaussian, i.e. $G_1 \cdot G_2 = G_3$, or

$$e^{-\alpha_1 r_A^2} e^{-\alpha_2 r_B^2} = K e^{-\gamma r_P^2}. \quad (2.30)$$

Expand Eqn (2.30) using the definition of $\mathbf{r}_A, \mathbf{r}_B, \mathbf{r}_P$ given above.

$$\begin{aligned}
e^{-\alpha_1 r_A^2 - \alpha_2 r_B^2} &= \exp[-(\alpha_1 + \alpha_2)\mathbf{r} \cdot \mathbf{r} + 2(\alpha_1 \mathbf{A} + \alpha_2 \mathbf{B}) \cdot \mathbf{r} \\
&\quad - \alpha_1 \mathbf{A} \cdot \mathbf{A} - \alpha_2 \mathbf{B} \cdot \mathbf{B}] \quad (2.31) \\
&= K \exp[-\gamma(\mathbf{r} \cdot \mathbf{r} - \mathbf{r} \cdot \mathbf{P} + \mathbf{P} \cdot \mathbf{P})] \quad (2.32)
\end{aligned}$$

Comparing terms,

$$\begin{aligned}
\gamma &= \alpha_1 + \alpha_2 \\
\gamma \mathbf{P} &= (\alpha_1 \mathbf{A} + \alpha_2 \mathbf{B}), \text{ thus } \quad \mathbf{P} = \frac{\alpha_1 \mathbf{A} + \alpha_2 \mathbf{B}}{\gamma} \quad (2.33)
\end{aligned}$$

which leads to the conclusion that

$$\begin{aligned}
K e^{-\gamma \mathbf{P} \cdot \mathbf{P}} &= e^{-\alpha_1 \mathbf{A} \cdot \mathbf{A} - \alpha_2 \mathbf{B} \cdot \mathbf{B}} \quad (2.34) \\
K &= e^{-\alpha_1 \mathbf{A} \cdot \mathbf{A} - \alpha_2 \mathbf{B} \cdot \mathbf{B} + \gamma \mathbf{P} \cdot \mathbf{P}} \quad (2.35)
\end{aligned}$$

From Eqn (2.33), we expand $\mathbf{P} \cdot \mathbf{P}$ and use that to get a final expression for $K$,

$$\begin{aligned}
\gamma \mathbf{P} \cdot \mathbf{P} &= \gamma^{-1} \left[ \alpha_1^2 \mathbf{A} \cdot \mathbf{A} + 2\alpha_1 \alpha_2 \mathbf{A} \cdot \mathbf{B} + \alpha_2^2 \mathbf{B} \cdot \mathbf{B} \right] \quad (2.36) \\
K &= \exp\left[ -\alpha_1 \mathbf{A} \cdot \mathbf{A} - \alpha_2 \mathbf{B} \cdot \mathbf{B} + (\alpha_1^2 \mathbf{A} \cdot \mathbf{A} + 2\alpha_1 \alpha_2 \mathbf{A} \cdot \mathbf{B} + \alpha_2^2 \mathbf{B} \cdot \mathbf{B})/\gamma \right] \\
&= \exp\left[ (-\alpha_1^2 \mathbf{A} \cdot \mathbf{A} - \alpha_1 \alpha_2 \mathbf{A} \cdot \mathbf{A} - \alpha_1 \alpha_2 \mathbf{B} \cdot \mathbf{B} - \alpha_2^2 \mathbf{B} \cdot \mathbf{B} \right. \\
&\quad \left. + \alpha_1^2 \mathbf{A} \cdot \mathbf{A} + 2\alpha_1 \alpha_2 \mathbf{A} \cdot \mathbf{B} + \alpha_2^2 \mathbf{B} \cdot \mathbf{B})\gamma^{-1} \right] \\
&= e^{-[\alpha_1 \alpha_2 (\overline{\mathbf{AB}}^2)/\gamma]} \quad (2.37)
\end{aligned}$$

if we define $\overline{\mathbf{AB}} = (\mathbf{A} - \mathbf{B})$. For two $s$-type functions,

$$e^{-\alpha_1 r_A^2} e^{-\alpha_2 r_B^2} = \exp\left[ -\alpha_1 \alpha_2 (\overline{\mathbf{AB}}^2)/\gamma \right] \exp\left[ -\gamma (\mathbf{r} - \mathbf{P})^2 \right] \quad (2.38)$$

For more general Cartesian Gaussians, ones with arbitrary angular momentum,

$$G_1 G_2 = x_A^{l_1} x_B^{l_2} y_A^{m_1} y_B^{m_2} z_A^{n_1} z_B^{n_2} \underbrace{e^{-(\alpha_1 \alpha_2 (\overline{\mathbf{AB}})^2/\gamma)}}_{K} e^{\gamma r_P^2} \quad (2.39)$$

where we've used Eqn (2.38) to take care of the product of the exponentials. Now, $x_A^{l_1}, x_B^{l_2}$ and the like need to be considered.

$$x_A^{l_1} x_B^{l_2} = (x - A_x)^{l_1}(x - B_x)^{l_2} \tag{2.40}$$

$$(x - A_x)^{l_1} = [(x - P_x) + (P_x - A_x)]^{l_1} = (x_P - (\overline{\mathbf{PA}})_x)^{l_1}. \tag{2.41}$$

Using a standard binomial expansion,

$$(x_P - (\overline{\mathbf{PA}})_x)^{l_1} = \sum_{i=0}^{l_1}(x_P)^i(\overline{\mathbf{PA}})_x^{l_1-i}\frac{l_1!}{i!(l_1-i)!} = \sum_{i=0}^{l_1}(x_P)^i(\overline{\mathbf{PA}})_x^{l_1-i}\binom{l_1}{i} \tag{2.42}$$

Likewise,

$$(x - B_x)^{l_2} = (x_P - (\overline{\mathbf{PB}})_x)^{l_2} = \sum_{j=0}^{l_2}(x_P)^j(\overline{\mathbf{PB}})_x^{l_2-j}\binom{l_2}{j}. \tag{2.43}$$

Using these, we can write $x_A^{l_1} x_B^{l_2}$ as a summation of $x_P$ to various powers.

$$x_A^{l_1} x_B^{l_2} = \sum_{k=0}^{l_1+l_2} x_P^k f_k(l_1, l_2, (\overline{\mathbf{PA}})_x, (\overline{\mathbf{PB}})_x). \tag{2.44}$$

The coefficient of $x_P^k$ in the product $x_A^{l_1} x_B^{l_2}$ is given by

$$f_k(l_1, l_2, \overline{\mathbf{PA}}_x, \overline{\mathbf{PB}}_x) = \sum_{i=0, l_1}^{i+j=k} \sum_{j=0, l_2} (\overline{\mathbf{PA}})_x^{l_1-i}\binom{l_1}{i}(\overline{\mathbf{PB}})_x^{l_2-j}\binom{l_2}{j} \tag{2.45}$$

Perhaps more conveniently for implementing in a computational scheme, the constrained double sum in the above expression for $f_k$ can be redefined as a single sum.

$$f_k = \sum_{q=\max(-k,k-2l_2)}^{\min(k,2l_1-k)^*} \binom{l_1}{i}\binom{l_2}{j}(\overline{\mathbf{PA}})_x^{l_1-i}(\overline{\mathbf{PB}})_x^{l_2-j} \tag{2.46}$$

$$2i = k + q$$
$$2j = k - q$$
$$\text{*increments of 2}$$

Whence we write the full Gaussian Product Theorem as Eqn (2.26). The derivation of Eqn (2.46) is left to the reader.

# 3  $S_{ij}$ – Overlap Integrals

## 3.1  Overlap of primitive $s$–functions on different centers

The integral we need to evaluate here is

$$\int \phi_1^*(\mathbf{r})\phi_2(\mathbf{r})d\mathbf{r} = \int e^{-\alpha_1 \mathbf{r}_A^2} e^{-\alpha_2 \mathbf{r}_B^2} d\mathbf{r} \tag{3.1}$$

Using the Gaussian Product Theorem as it appears in Eqn (2.38)

$$S_{12} = \int e^{-\alpha_1 \alpha_2 (\overline{\mathbf{AB}})^2/\gamma} e^{-\gamma r_P^2} d\mathbf{r} \tag{3.2}$$

$$= e^{-\alpha_1 \alpha_2 (\overline{\mathbf{AB}})^2/\gamma} \int_{-\infty}^{\infty} e^{-\gamma x_P^2} dx \int_{-\infty}^{\infty} e^{-\gamma y_P^2} dy \int_{-\infty}^{\infty} e^{-\gamma z_P^2} dz \tag{3.3}$$

$$S_{12} = e^{-\alpha_1 \alpha_2 (\overline{\mathbf{AB}})^2/\gamma} \left(\frac{\pi}{\gamma}\right)^{3/2} \tag{3.4}$$

## 3.2  Overlap of contracted $s$–functions

Take now $\phi_1(\mathbf{r})$ to be centered on $\mathbf{A}$ and $\phi_2(\mathbf{r})$ to be centered on $\mathbf{B}$, as

$$\phi_1(\mathbf{r}) = N_1 \sum_i^n a_i e^{-\alpha_i r_A^2}, \phi_2(\mathbf{r}) = N_2 \sum_j^m b_j e^{-\beta_j r_B^2} \tag{3.5}$$

$$S_{12} = \int \phi_1^*(\mathbf{r})\phi_2(\mathbf{r})d\mathbf{r} = N_1 N_2 \sum_i^n \sum_j^m a_i b_j \int e^{-\alpha_i r_A^2} e^{-\beta_j r_B^2} d\mathbf{r} \tag{3.6}$$

Examining one term in the double sum,

$$\int e^{-\alpha_i \mathbf{r}_A^2} e^{-\beta_j \mathbf{r}_B^2} d\mathbf{r} = \int e^{-\alpha_i \beta_j (\overline{\mathbf{AB}})^2/\gamma} e^{-\gamma r_P^2} d\mathbf{r} \tag{3.7}$$

$$= e^{-\alpha_i \beta_j (\overline{\mathbf{AB}})^2/\gamma_{ij}} \left(\frac{\pi}{\gamma_{ij}}\right)^{3/2} \tag{3.8}$$

where $\gamma_{ij} = \alpha_i + \beta_j$ and $\mathbf{P}_{ij} = \frac{\alpha_i \mathbf{A} + \beta_j \mathbf{B}}{\gamma_{ij}}$. So

$$S_{12} = N_1 N_2 \sum_i^n \sum_j^m a_i b_j e^{-\alpha_i \beta_j (\overline{\mathbf{AB}})^2 / \gamma_{ij}} \left[\frac{\pi}{\gamma_{ij}}\right]^{3/2} \quad (3.9)$$

## 3.3 Overlap of primitive arbitrary angular momentum functions

Overlap of arbitrary–$l$ functions:

$$S_{12} = \int G_1(\alpha_1, \mathbf{A}, l_1, m_1, n_1) G_2(\alpha_2, \mathbf{B}, l_2, m_2, n_2) d\mathbf{r} \quad (3.10)$$

$$= \int x_A^{l_1} x_B^{l_2} y_A^{m_1} y_B^{m_2} z_A^{n_1} z_B^{n_2} \exp[-\alpha_1 \alpha_2 (\overline{\mathbf{AB}})^2 / \gamma] e^{-\gamma x_P^2} e^{-\gamma y_P^2} e^{-\gamma z_P^2} \quad (3.11)$$

with $\gamma$ and $\mathbf{P}$ defined as before. Applying the fullness of the GPT [Eqn (2.26)],

$$S_{12} = \exp[-\alpha_1 \alpha_2 (\overline{\mathbf{AB}})^2 / \gamma] I_x I_y I_z. \quad (3.12)$$

where

$$I_x = \int \sum_{i=0}^{l_1 + l_2} f_i(l_1, l_2, \overline{\mathbf{PA}}_x, \overline{\mathbf{PB}}_x) x_P^i e^{-\gamma x_P^2} dx \quad (3.13)$$

$$= \sum_{i=0}^{l_1 + l_2} f_i(l_1, l_2, \overline{\mathbf{PA}}_x, \overline{\mathbf{PB}}_x) \int_{-\infty}^{\infty} x_P^i e^{-\gamma x_P^2} dx \quad (3.14)$$

Noting that any odd value of $i$ produces a zero integral, and then using Eqn (2.22) for $\int x_P^i e^{-\gamma x_P^2} dx$, we finally write $I_x$ as

$$I_x = \sum_{i=0}^{(l_1 + l_2)/2} f_{2i}(l_1, l_2, \overline{\mathbf{PA}}_x, PB_x) \frac{(2i-1)!!}{(2\gamma)^i} \left(\frac{\pi}{\gamma}\right)^{1/2}. \quad (3.15)$$

The last remaining case is the overlap of two contracted Gaussians of arbitrary angular momentum is left out of the consideration. It is not difficult to write simple routines to compute overlap integrals over primitive Gaussian functions of arbitrary angular momentum using Eqns (3.12) and (3.15) and use those routines in the evaluation of overlap integrals over contracted functions.

## 4  $T_{ij}$–Kinetic Energy Integrals

The kinetic energy operator is $-\frac{1}{2}\nabla^2$, or $-\frac{1}{2}(\partial^2/\partial x^2 + \partial^2/\partial y^2 + \partial^2/\partial z^2)$ in Cartesian coordinates. So the kinetic energy integral over general, uncontracted Gaussian functions is

$$\begin{aligned}
\mathbf{T}_{12} &= \int \phi_1^*(\mathbf{r})(-\frac{1}{2}\nabla^2)\phi_2 d\mathbf{r} \\
&= -\frac{1}{2}\int x_A^{l_1} y_A^{m_1} z_A^{n_1} e^{-\alpha_1 r_A^2}(\frac{\partial^2}{\partial x^2} + \frac{\partial^2}{\partial y^2} + \frac{\partial^2}{\partial z^2}) x_B^{l_2} y_B^{m_2} z_B^{n_2} e^{-\alpha_1 r_A^2} d\mathbf{r} \quad (4.1) \\
&= I_x + I_y + I_z
\end{aligned}$$

where we now define $I_x$ as

$$I_x = -\frac{1}{2}\int x_A^{l_1} y_A^{m_1} z_A^{n_1} e^{-\alpha_1 r_A^2}(\frac{\partial^2}{\partial x^2}) x_B^{l_2} y_B^{m_2} z_B^{n_2} e^{-\alpha_1 r_A^2} d\mathbf{r} \quad (4.2)$$

Now we need to determine the action of the Lagrangian (or any piece thereof) on a particular Gaussian function. Sequentially applying the differential operator,

$$\frac{\partial}{\partial x}(x_B^{l_2} e^{-\alpha_2 x_B^2}) = l_2 x_B^{l_2-1} e^{-\alpha_2 x_B^2} - 2\alpha_2 x_B^{l_2+1} e^{-\alpha_2 x_B^2} \quad (4.3)$$

$$\begin{aligned}
\frac{\partial}{\partial x}(\frac{\partial}{\partial x}(x_B^{l_2} e^{-\alpha_2 x_B^2})) &= l_2(l_2-1)x_B^{l_2-2} e^{-\alpha_2 x_B^2} - 2\alpha_2(2l_2+1)x_B^{l_2} e^{-\alpha_2 x_B^2} \\
&\quad + 4\alpha_2^2 x_B^{l_2+2} e^{-\alpha_2 x_B^2} \quad (4.4)
\end{aligned}$$

$$\begin{aligned}
-\frac{1}{2}\frac{\partial^2}{\partial x^2}(x_b^{l_2} e^{-\alpha_2 x_B^2}) &= -\frac{l_2(l_2-1)}{2} x_B^{l_2-2} e^{-\alpha_2 x_B^2} \\
&\quad + \alpha_2(2l_2+1)x_B^{l_2} e^{-\alpha_2 x_B^2} - 2\alpha_2^2 x_B^{l_2+2} e^{-\alpha_2 x_B^2} \quad (4.5)
\end{aligned}$$

Clearly, this is just a sum of three Gaussian functions related to the original by a shift of 0, 2, or -2 in the angular momentum portion, aside from some constants.

## 4.1 Asymmetric form of $T_{ij}$

Simply applying the results shown in Eqn (4.5) within Eqn (4.1) gives a form of $T_{ij}$ which appears as a sum of three overlap–type integrals with various multiplicative constants. To display the particular overlap integrals involved in that sum we will use a particular notation derived from the bra and ket notation common in physics. Let $\langle \pm n|_\gamma$ denote a Gaussian where the angular momentum has been increased or decreased by $n$ in the $\gamma$ coordinate. In other words,
$$\langle +2|_x = x^{l+2} y^m z^n e^{-\alpha r^2} \qquad (4.6)$$
Thus, given that the overlap between two Gaussians $G_1$ and $G_2$ is
$$\int G_1 G_2 = \langle 0|0\rangle, \qquad (4.7)$$
the construction $\langle 0| + 2\rangle_x$ denotes an overlap integral between $G_1$ and a Gaussian derived from $G_2$ by incrementing the exponent of $x$ by 2. In this way, we can write the asymmetric form of the kinetic energy integral using Eqns (4.2) and (4.5) as
$$I_x = \alpha_2(2l_2 + 1)\langle 0|0\rangle - 2\alpha_2^2 \langle 0| + 2\rangle_x - \frac{l_2(l_2 - 1)}{2}\langle 0| - 2\rangle_x \qquad (4.8)$$

## 4.2 Symmetric form of $T_{ij}$

Time to try a different approach. Starting with the old definition of $I_x$,
$$I_x = -\frac{1}{2}\iiint \phi_1^*(\mathbf{r})\frac{\partial^2}{\partial x^2}\phi_2(\mathbf{r})dxdydz \qquad (4.9)$$
and integrating by parts in $x$,
$$I_x = -\frac{1}{2}\left[\iint \left(\phi_1^*(\mathbf{r})\frac{\partial \phi_2(\mathbf{r})}{\partial x}\right)\Big|_{-\infty}^{+\infty} dydz - \iiint \frac{\partial \phi_1^*(\mathbf{r})}{\partial x}\frac{\partial \phi_2(\mathbf{r})}{\partial x}dxdydz\right] \qquad (4.10)$$

The first term is of course zero because both $\phi_1(\mathbf{r})$ and $\partial\phi_2(\mathbf{r})/\partial x$ go to zero as $x \to \pm\infty$. So

$$I_x = \frac{1}{2} \iiint \frac{\partial \phi_1}{\partial x} \frac{\partial \phi_2}{\partial x} dx dy dz \qquad (4.11)$$

Recalling Eqn (4.3),

$$\begin{aligned} I_x &= \frac{1}{2} \iiint \left[ l_2 x_A^{l_1-1} - 2\alpha_1 x_A^{l_1+1} \right] y_A^{m_1} z_A^{n_1} e^{\alpha_1 r_A^2} \\ &\quad \cdot \left[ l_2 x_B^{l_2-1} - 2\alpha_2 x_B^{l_2+1} \right] y_B^{m_2} z_B^{n_2} e^{-\alpha_2 r_B^2} dx dy dz. \end{aligned} \qquad (4.12)$$

Thus we can reduce this cumbersome notation to something a little more convenient to code up.

$$\begin{aligned} I_x &= \frac{1}{2} l_1 l_2 \langle -1| -1 \rangle_x + 2\alpha_1 \alpha_2 \langle +1| +1 \rangle_x \\ &\quad - \alpha_1 l_2 \langle +1| -1 \rangle_x - \alpha_2 l_1 \langle -1| +1 \rangle_x \end{aligned} \qquad (4.13)$$

It is somewhat more appealing, since $T_{ij}$ should be a symmetric matrix, i.e. $T_{ij} = T_{ji}$. This is an obvious truth when Eqn (4.13) is used to calculate T, but is not so from the asymmetric form. Good news – both expressions give exactly the same result.

# 5  $V_{ij}$ – Nuclear Attraction Integrals

## 5.1  The need for a transformation

Since the potential energy is due to Coulomb interaction of the nuclei with the electron in question, the operator to deal with is $\frac{1}{r_C}$, where $r_C = |\mathbf{r} - \mathbf{C}|$. Thus the integral we need to evaluate is

$$\begin{aligned}
V_{ij}^C &= \int \phi_i^*(\mathbf{r}) \frac{1}{r_C} \phi_j(\mathbf{r}) d\mathbf{r} \\
&= \int x_A^{l_1} y_A^{m_1} z_A^{n_1} e^{-\alpha_1 r_A^2} \frac{1}{r_C} x_B^{l_2} y_B^{m_2} z_B^{n_2} e^{-\alpha_2 r_B^2} d\mathbf{r}
\end{aligned} \quad (5.1)$$

Since the operator does not affect the operand ($\phi$), we can combine the two Gaussian functions via the Gaussian product theorem, and make the final statement

$$\begin{aligned}
V_{ij}^C &= K \sum_l \sum_m \sum_n f_l(l_1, l_2, \overline{\mathbf{PA}}_x, \overline{\mathbf{PB}}_x) f_m(m_1, m_2, \overline{\mathbf{PA}}_y, \overline{\mathbf{PB}}_y) \\
&\cdot f_n(n_1, n_2, \overline{\mathbf{PA}}_z, \overline{\mathbf{PB}}_z) \int x_P^l y_P^m z_P^n e^{-\gamma r_P^2} \frac{1}{r_C} d\mathbf{r}
\end{aligned} \quad (5.2)$$

where $K = e^{-\alpha_1 \alpha_2 (\overline{\mathbf{AB}}^2/\gamma)}$. This is still intractable. Indeed, what do we have under the integral sign?[#] The usual Gaussian-like term $x_P^l y_P^m z_P^n e^{-\gamma r_P^2}$ is not the bottleneck, it is easily representable as a product of three terms each depending on $x$, $y$, and $z$ respectively. The term that prevents us from separating the three-dimensional integral over $\mathbf{r}$ into three one-dimensional integrals is $\frac{1}{r_C}$. Recall that

$$r_C = \sqrt{x_C^2 + y_C^2 + z_C^2} \quad (5.3)$$

At this point, we want to apply some sort of transform to the $\frac{1}{r_C}$ to turn it into some sort of an exponential which can be combined with the other Gaussians and result in resolution of the variables. There are many possibilities – the most commonly used are Laplace or Fourier transforms. Let's look at the Laplace transform solution in detail.

---

[#] Remember, it's a three-dimensional integral, **r** is a vector.

## 5.2 Laplace transform

Use the standard Laplace transform,

$$r^{-\lambda} = \left[\Gamma(\frac{\lambda}{2})\right]^{-1} \int_0^\infty e^{-sr^2} s^{\lambda/2-1} ds, \tag{5.4}$$

where $\Gamma(x)$ is the standard gamma function. You can just evaluate the RHS to confirm this. We want the instance where $\lambda = 1$, thus

$$\frac{1}{r_C} = \pi^{-1/2} \int_0^\infty e^{-sr_C^2} s^{-1/2} ds. \tag{5.5}$$

What occurs when we use this in the context of a potential energy integral involving only $s$–functions?

$$\begin{aligned} V &= \int e^{-\alpha_1 r_A^2} e^{-\alpha_2 r_B^2} \frac{1}{r_C} d\mathbf{r} \\ &= K \int e^{-\gamma r_P^2} \frac{1}{r_C} d\mathbf{r} \\ &= K\pi^{-1/2} \int e^{-\gamma r_P^2} \int_0^\infty e^{-sr_C^2} s^{-1/2} ds d\mathbf{r} \end{aligned} \tag{5.6}$$

Conveniently, the Laplace transform takes $r_C^{-1}$ into a function with the appearance of an $s$-type Gaussian of orbital exponent $s$ centered at $\mathbf{C}$. A second application of the GPT and we can switch the order of integration, evaluating the integral over $s$ second.

$$\begin{align}
V &= K\pi^{-1/2} \int_0^\infty ds\, s^{-1/2} \int e^{-\gamma r_P^2} e^{-s r_C^2} d\mathbf{r} \\
&= K\pi^{-1/2} \int_0^\infty ds\, s^{-1/2} e^{-\gamma s \overline{\mathbf{PC}}^2/(\gamma+s)} \int d\mathbf{r}\, e^{-(\gamma+s) r_D^2} \tag{5.7} \\
&= K\pi \int_0^\infty ds\, s^{-1/2} (\gamma+s)^{-3/2} e^{-\gamma s \overline{\mathbf{PC}}^2/(\gamma+s)}. \tag{5.8}
\end{align}$$

Now making the substitution $t^2 = \frac{s}{(\gamma+s)}$, $ds = \frac{2}{\gamma} s^{1/2} (\gamma+s)^{3/2} dt$ amazingly cancels just about everything leaving

$$V = \frac{2K\pi}{\gamma} \int_0^1 e^{-\gamma \overline{\mathbf{PC}}^2 t^2} dt \tag{5.9}$$

which can be rewritten in terms of a standard error function when $\mathbf{P} \neq \mathbf{C}$,

$$\operatorname{erf}(x) = \frac{2}{\pi^{1/2}} \int_0^x e^{-t^2} dt \tag{5.10}$$

$$V = \frac{K\pi^{3/2}}{\gamma^{3/2} \overline{\mathbf{PC}}} \operatorname{erf}(\gamma^{1/2} \overline{\mathbf{PC}}),\ \mathbf{P} \neq \mathbf{C} \tag{5.11}$$

or as

$$V = \frac{2K\pi}{\gamma},\ \mathbf{P} = \mathbf{C} \tag{5.12}$$

The case of arbitrary angular momentum Gaussians is done likewise, only instead of the standard error function we get error function-like integrals referred to as an *incomplete gamma function*:

$$F_m(T) = \int_0^1 t^{2m} e^{-T t^2} dt \tag{5.13}$$

Note that

$$F_0(T) = \frac{\pi^{1/2}}{2\sqrt{T}}\mathrm{erf}(\sqrt{T}), T > 0 \quad (5.14)$$
$$F_0(0) = 1 \quad (5.15)$$

## 5.3 Fourier transform

The Fourier transform approach is essentially very similar, the only difference is that $\frac{1}{r_C}$ is represented as a *three*-dimensional integral:

$$\frac{1}{r_C} = \frac{1}{2\pi^2}\iiint k^{-2}e^{i\mathbf{k}\cdot\mathbf{r}_C}d\mathbf{k} \quad (5.16)$$

A substitution of Eqn (5.16) into Eqn (5.2) and a reversal of the order of integration allow the integral over $\mathbf{r}$ to be separated into three one-dimensional integrals. The resulting derivation is a little bit longer than in the case of Laplace transform, and the final expression looks a bit different, but both formulae give the same values, and that's what counts.

# 6 Electron Repulsion Integrals

The electron repulsion integral (ERI) is

$$\langle ik|jl\rangle = (ij|kl) = \int \phi_i^*(\mathbf{r}_1)\phi_k^*(\mathbf{r}_2)\frac{1}{r_{12}}\phi_j(\mathbf{r}_1)\phi_l(\mathbf{r}_2)d\mathbf{r}_1 d\mathbf{r}_2 \qquad (6.1)$$

ERIs are two-electron integrals as opposed to the one-electron integrals that we've dealt with so far. Nevertheless, the techniques that we have described in the previous sections can be applied to evaluate ERIs in closed form. We are going to sketch the strategy of a computation and present the final expression for an ERI over four arbitrary angular momentum primitives centered on four different centers.

1. We begin with the following general primitive ERI:

$$\begin{aligned}
\phi_i(\mathbf{r}_1) &= x_{1A}^{l_a}y_{1A}^{m_a}z_{1A}^{n_a}\exp(-\alpha_1 r_{1A}^2) \\
\phi_j(\mathbf{r}_1) &= x_{1B}^{l_b}y_{1B}^{m_b}z_{1B}^{n_b}\exp(-\alpha_2 r_{1B}^2) \\
\phi_k(\mathbf{r}_2) &= x_{2C}^{l_c}y_{2C}^{m_c}z_{2C}^{n_c}\exp(-\alpha_3 r_{2C}^2) \\
\phi_l(\mathbf{r}_2) &= x_{2D}^{l_d}y_{2D}^{m_d}z_{2D}^{n_d}\exp(-\alpha_4 r_{2D}^2) \\
I &= \int \phi_i(\mathbf{r}_1)\phi_j(\mathbf{r}_1)\frac{1}{r_{12}}\phi_k(\mathbf{r}_2)\phi_l(\mathbf{r}_2)d\mathbf{r}_1 d\mathbf{r}_2 \qquad (6.2)
\end{aligned}$$

2. Combine $\phi_i$ with $\phi_j$ and $\phi_k$ with $\phi_l$ using the GPT.

3. The $\frac{1}{r_{12}}$ factor in the integral demands some sort of integral transform to be applied to it, just like in the case of the nuclear attraction integrals:

$$\frac{1}{r_{12}} = \frac{1}{2\pi^2}\iiint k^{-2}e^{i\mathbf{k}\cdot\mathbf{r}_{12}}d\mathbf{k} \qquad (6.3)$$

4. Insertion of Eqn (6.3) into the composite expression we've obtained in the first step yields a nine-dimensional integral which might look nastier than the six-dimensional ERI (6.2) we began with. However, we can switch the order of integration, and evaluate the integrals over six spatial coordinates ($x_1$, $x_2$, $y_1$, etc.).

5. The resulting expression is a three dimensional integral over variables $k_x$, $k_y$, and $k_z$, which can be evaluated quite easily. The resulting monstrous 15-fold summation formula looks like this:

$$I = \frac{2\pi^{5/2} K_1 K_2}{\gamma_p \gamma_q (\gamma_p + \gamma_q)^{1/2}} \sum_{l_p,l_q} \sum_{u_1,u_2} \sum_{t'} G(x) \sum_{m_p,m_q} \sum_{v_p,v_q}$$
$$\sum_{t''} G(y) \sum_{n_p,n_q} \sum_{\omega_1,\omega_2} \sum_{t'''} G(z) F_\zeta(\overline{\mathbf{PQ}}^2 \gamma_p \gamma_q / (\gamma_p + \gamma_q)). \quad (6.4)$$

where

$$\zeta = l_p + l_q + n_p + n_q + m_p + m_q - 2u_1 - 2u_2 - 2v_1 - 2v_2 -$$
$$2w_1 - 2w_2 - t' - t'' - t'''$$
$$K_1 = \exp(-\alpha_1 \alpha_2 \overline{\mathbf{AB}}^2 / \gamma_p)$$
$$K_2 = \exp(-\alpha_3 \alpha_4 \overline{\mathbf{CD}}^2 / \gamma_q)$$
$$G(x) = (-1)^{l_p + t'} f_{l_p}(l_a, l_b, PA_x, PB_x) f_{l_q}(l_c, l_d, QC_x, QD_x)$$
$$\times (PQ_x)^{l_p + l_q - 2u_1 - 2u_2 - 2t'} \left(\frac{\gamma_p \gamma_q}{\gamma_p + \gamma_q}\right)^{l_p + l_q - 2u_1 - 2u_2 - t'}$$
$$\times \frac{(\gamma_p)^{u_1 - l_p} (\gamma_q)^{u_2 - l_q} l_p! l_q! (l_p + l_q - 2u_1 - 2u_2)! 4^{-u_1 - u_2 - t'}}{u_1! u_2! (l_p - 2u_1)! (l_q - 2u_2)! (l_p + l_q - 2u_1 - 2u_2 - 2t')! (t')!}$$
$$\mathrm{F}_m(T) = \int_0^1 u^{2m} \exp(-Tu^2) du$$

$$0 \le l_p \le l_a + l_b \quad 0 \le l_q \le l_c + l_d$$
$$0 \le u_1 \le l_p/2 \quad 0 \le u_2 \le l_q/2$$
$$0 \le t' \le (l_p + l_q - 2u_1 - 2u_2)/2$$

etc.

In the derivation we used the Fourier transformation of the $\frac{1}{r_{12}}$ factor, but, likewise, the standard Laplace or Gaussian transforms may be used.

What can we learn from Eqn (6.4)?

- We can immediately obtain a compact expression for an ERI over 4 primitive $s$-type Gaussians:

$$(ss|ss) = \frac{2\pi^{5/2} K_1 K_2}{\gamma_p \gamma_q (\gamma_p + \gamma_q)^{1/2}} F_0(\overline{\mathbf{PQ}}^2 \gamma_p \gamma_q / (\gamma_p + \gamma_q)) \quad (6.5)$$

- Computing ERI of any but $(ss|ss)$ type in closed form is not very practical. Coding it up is relatively easy, but the efficiency is very poor, especially if you need to compute more than just one integral.

- Eqn (6.4) may be rewritten as

$$I = \sum_{m=0}^{L} C_m F_m(T) \quad (6.6)$$

where $L = l_a + m_a + n_a + l_b + \cdots + n_d$; $T$ is $\overline{\mathbf{PQ}}^2 \gamma_p \gamma_q / (\gamma_p + \gamma_q)$ and thus independent of $l_a$, $m_a$, $n_a$, $l_b$, etc. [#] "Angular" coefficients $C_m$ in turn do depend on the angular momentum indices $l_a$, etc. It turns out that Eqn (6.6) is very general and serves as a starting point for many ERI evaluation methods.

Having said this, let's look at more practical methods of computing ERIs.

---

[#] Let's refer to the exponents of $x$, $y$, and $z$ as "angular momentum indices"

# 7 Practical methods of computing the electron repulsion integrals

Up to this point we've used a notation system that we feel is the most suitable for manipulating closed form expressions for one-electron integrals. To move on further we have to enhance our notation to bring it in accordance with the one used in the literature. Following Obara and Saika [2], we write the unnormalized Cartesian Gaussian function centered at $\mathbf{R}$ as

$$\begin{aligned}\phi(\mathbf{r};\zeta,\mathbf{n},\mathbf{R}) &= (x-R_x)^{n_x}(y-R_y)^{n_y}(z-R_z)^{n_z}\\&\quad \times \exp[-\zeta(\mathbf{r}-\mathbf{R})^2]\,,\end{aligned} \qquad (7.1)$$

where $\mathbf{r}$ is the coordinate vector of the electron, $\zeta$ is the orbital exponent, and $\mathbf{n}$ is a set of non-negative integers. Sum of $n_x$, $n_y$, and $n_z$ will be denoted $\lambda$ and is the angular momentum or orbital quantum number. Hereafter $\mathbf{n}$ will be termed the angular momentum index. Henceforth, $n_i$ will refer to the $i$-th component of $\mathbf{n}$, where $i \in \{x,y,z\}$. Basic vector addition rules will apply to these vector-like triads of numbers, e.g. $\mathbf{n}+\mathbf{1}_x \equiv \{n_x+1, n_y, n_z\}$. A set of $(\lambda+1)(\lambda+2)/2$ functions with the same $\lambda$, $\zeta$, and $\mathbf{R}$ but different $\mathbf{n}$ form a *Cartesian shell*, or just a *shell*. A set of integrals $\{(\mathbf{ab}|\mathbf{cd})\}$ over all possible combinations of functions $\mathbf{a} \in$ ShellA, $\mathbf{b} \in$ ShellB, etc. is termed a *shell*, or *quartet*, or *class* of integrals. For example, a $(ps|sd)$ class consists of $3 \times 1 \times 1 \times 6 = 18$ integrals.

The last comment before we dive in. We feel that it is beyond our ability to give here an exhaustive review of all available methods with all the details so that an unprepared person can read this document, understand thoroughly every concept, and be able to apply them in practice (read – write a computer code). Instead, we will try to give an overview of key ideas and most important algorithms and refer the reader to other, more informative and thorough sources.

## 7.1 Numerical integration methods

The first group of methods evaluates ERIs by numerical integration similar to that used for integrating functions of one variable in college calculus. The idea of numerical one-dimensional integration is to approximate an integral by a finite sum:

$$\int_a^b f(x)dx = \sum_i^N f(x_i)w_i \; ; \quad x_i \in [a;b] \tag{7.2}$$

Of course, we want to compute a *six − dimensional* integral, and our approach will have to be a little bit different (why cannot we just use a formula analogous to Eqn (7.2)? Read the answer at the bottom of this page).[#] The trick is to represent an ERI as a one-dimensional integral over some function. Let's start with Eqn (6.6). After plugging the definition of the incomplete gamma function in and moving the summation sign inside the integral we obtain

$$\begin{aligned} I &= \int_0^1 \sum_{m=0}^L C_m t^{2m} \exp(-Tt^2)dt \\ &= \int_0^1 P_L(t) \exp(-Tt^2)dt \; , \end{aligned} \tag{7.3}$$

where $P_L(t)$ is an even polynomial of degree $2L$ with coefficients $C_m$.

Eqn (7.3) has the following form:

$$I = \int_a^b K_l(x)w(x)dx \tag{7.4}$$

where $K_l(x)$ is a polynomial of degree $l$, $w(x)$ is positive on the interval $[a,b]$. The theory of orthogonal polynomials offers a way of computing this type of integrals *exactly*. Let a set of polynomials $\{S_n(x)\}$ be defined on $[a; b]$ and orthogonal in the sense that

$$\int_a^b S_n(x) S_m(x) w(x) dx = \delta_{mn} , \tag{7.5}$$

then integral (7.4) can be evaluated exactly by an $n$-point numerical quadrature formula:

$$I = \sum_{i=1}^n K_l(t_i) W_i , \tag{7.6}$$

where $n \geq l$, $t_i$ is the $i$-th positive zero of $S_n(x)$, and $W_i$ is the corresponding weight factor computed as

$$W_i = \int_a^b L_i(x) w(x) dx , \tag{7.7}$$

where $L_i(x)$ is the Lagrange polynomial:

$$L_i(x) = \frac{(x - t_1) \cdots (x - t_{i-1})(x - t_{i+1}) \cdots (x - t_n)}{(t_i - t_1) \cdots (t_i - t_{i-1})(t_i - t_{i+1}) \cdots (t_i - t_n)} \tag{7.8}$$

---

[#] Because the required six-fold summation would be computationally intractable

All the glorious details with the proofs can be found in any textbook on numerical methods.

Fortunately, there exists a set of polynomials $\{R_i(t, T)\}$, known as Rys polynomials, that satisfy all necessary conditions:

- $R_n(t, T)$ is an even polynomial of degree $2n$ in the variable $t$;
- For any real $T$ there exists an infinite set of such entities orthogonal to each other:

$$\int_0^1 R_n(t, T) R_m(t, T) \exp(-Tt^2) dt = \delta_{mn} \tag{7.9}$$

Hence, the ERI can be evaluated as

$$I = \sum_{i=1}^{n} P_L(t_i) W_i ,  \qquad (7.10)$$

where $n$ is an integer greater than $L/2$, $t_i$ is a positive zero of the $n$-th Rys polynomial

$$R_n(t_i, T) = 0 , \qquad (7.11)$$

and $W_i$ is the corresponding weight which depends on $T$. Since $W_i$ depends on $T$ only, and every integral in a given class of integrals will share the same set of $W_i$, it is beneficial to use Eqn (7.10) to compute whole classes of integrals rather than a single integral. In fact, all practical strategies of ERI evaluation employ this so-called shell structure of integrals.

What are potential advantages of using Eqn (7.10) instead of Eqn (6.6)? At first glance – none. To compute $P_L$ one has to know coefficients $C_m$, computing which is the main obstacle in using Eqn (6.6). Yet it turns out there exists a way to determine numerical values of $P_L(t_i)$ without computing $C_m$. The details are too lengthy to be presented here and can be found in a paper by Dupuis, Rys, and King [3]. The resulting expression is written as

$$I \sim \sum_{i=1}^{N} I_x(u_i) I_y(u_i) I_z(u_i) W_i , \qquad (7.12)$$

where $u_i \sim t_i/(1-t_i^2)^{1/2}$. Quantities $I_i$ depend on angular momentum indices and can be computed numerically or recursively.[4] Most recently, Ishida[5] have proposed several very efficient algorithms based on Eqn (7.12) and its variations, in which he computes $I_i$ numerically and recursively.

## 7.2 Recursive methods

"Recursive" implies computation of ERI from other ERIs of lower angular momentum. The first "recursive" method of computing ERIs was suggested by Boys. He applied the differential relation for Cartesian Gaussians

$$\frac{\partial}{\partial R_i} = 2\zeta \phi(\mathbf{r}, \zeta, \mathbf{n} + \mathbf{1}_i, \mathbf{R}) - n_i \phi(\mathbf{r}, \zeta, \mathbf{n} - \mathbf{1}_i, \mathbf{R}) \qquad (7.13)$$

to the expression for the $(ss|ss)$ integral given in Eqn. (6.5) to obtain the corresponding $(ps|ss)$ integral. Hence, multiple applications yield higher angular momentum ERIs.

McMurchie and Davidson proposed recurrence relations for ERIs over Hermite Gaussian functions #to compute the integrals over Cartesian Gaussians.[6] Their algorithm is a significant improvement over Boys' method, but it requires a special transformation of the resulting integrals over Hermite Gaussians back into the Cartesian basis.

Stagnant at the moment, the field was revitalized by Obara and Saika in 1986.[2] Obara and Saika also used the differential relation for Cartesian Gaussians (7.13). Their approach treats one-, two-, and potentially $n$-electron integrals on equal footing. OS suggested the following recurrence relation for

---

[#] Hermite Gaussians are a special form of Gaussian functions which resemble the eigenfunctions of the harmonic oscillator. Their use is motivated by extreme simplicity of the differential relations for this type of Gaussians. For more details see ...

ERIs over *primitive* Cartesian Gaussians: #

$$[\mathbf{a}+1_i, \mathbf{b}|\mathbf{cd}]^{(m)} = PA_i[\mathbf{ab}|\mathbf{cd}]^{(m)} + WP_i[\mathbf{ab}|\mathbf{cd}]^{(m+1)}$$
$$+ \frac{a_i}{2\zeta}\left([\mathbf{a}-1_i, \mathbf{b}|\mathbf{cd}]^{(m)} - \frac{\eta}{\zeta+\eta}[\mathbf{a}-1_i, \mathbf{b}|\mathbf{cd}]^{(m+1)}\right)$$
$$+ \frac{b_i}{2\zeta}\left([\mathbf{a}, \mathbf{b}-1_i|\mathbf{cd}]^{(m)} - \frac{\eta}{\zeta+\eta}[\mathbf{a}, \mathbf{b}-1_i|\mathbf{cd}]^{(m+1)}\right)$$
$$+ \frac{c_i}{2(\zeta+\eta)}[\mathbf{ab}|\mathbf{c}-1_i, \mathbf{d}]^{(m+1)} + \frac{d_i}{2(\zeta+\eta)}[\mathbf{ab}|\mathbf{c}, \mathbf{d}-1_i]^{(m+1)} \,, \quad (7.14)$$

where

$$(\mathbf{ab}|\mathbf{cd})^{(m)} = \frac{2}{\sqrt{\pi}} \int_0^\infty du \left(\frac{u^2}{\rho+u^2}\right)^m (\mathbf{ab}|u|\mathbf{cd}) \quad (7.15)$$

$$(\mathbf{ab}|u|\mathbf{cd}) = \int d\mathbf{r}_1 d\mathbf{r}_2 \phi(\mathbf{r}_1; \zeta_a, \mathbf{a}, \mathbf{A})\phi(\mathbf{r}_1; \zeta_b, \mathbf{b}, \mathbf{B})\exp(-u^2 r_{12}^2) \times$$
$$\phi(\mathbf{r}_2; \zeta_c, \mathbf{c}, \mathbf{C})\phi(\mathbf{r}_2; \zeta_d, \mathbf{d}, \mathbf{D}) \,, \quad (7.16)$$

and

$$\zeta = \zeta_a + \zeta_b \quad (7.17)$$
$$\eta = \zeta_c + \zeta_d \quad (7.18)$$
$$\rho = \frac{\zeta\eta}{\zeta+\eta} \quad (7.19)$$
$$\mathbf{P} = \frac{\zeta_a \mathbf{A} + \zeta_b \mathbf{B}}{\zeta_a + \zeta_b} \quad (7.20)$$
$$\mathbf{Q} = \frac{\zeta_c \mathbf{C} + \zeta_d \mathbf{D}}{\zeta_c + \zeta_d} \,. \quad (7.21)$$

$(\mathbf{ab}|\mathbf{cd})^{(m)}$ is an auxiliary ERI playing a central role in OS manipulations. Note that $(\mathbf{ab}|\mathbf{cd})^{(0)}$ is a "true" ERI.

---

# The presented equation is only one of the four possible cases, the other three being the relations increasing angular momentum on centers **B**, **C**, and **D**.

Thus, Eqn (7.14) allows to compute any ERI from a set of $(ss|ss)^m$ integrals

$$(00|00)^{(m)} \sim F_m(T) \qquad (7.22)$$

with $0 \leq m \leq \lambda_a + \lambda_b + \lambda_c + \lambda_d$. This result is of course equivalent to Eqn (6.6). However, painful evaluation of the coefficients in the sum (6.6) is replaced by a recursive application of a simple relation.

Of course, the OS RR allows the shell structure of ERIs to be exploited. The strategy of computing primitive ERIs using OS RR is simple:

1. compute all necessary $F_m(T)$;
2. apply Eqn (7.14) repeatedly to compute the target (shell) of integrals.

To compute contracted ERIs all possible combinations of primitives have to be evaluated separately using steps 1 and 2 and then contracted together to form the final value. Authors found a computer implementation of their algorithm to be vastly superior to other programs available at the time.

In an attempt to improve the performance of the OS RR for contracted ERIs, Head-Gordon and Pople[7] suggested two complementary relations. The first one is termed the Vertical Recurrence Relation (VRR) and is a special case of the OS RR with **b** and **d** set to 0:

$$[\mathbf{a} + 1_i, 0|\mathbf{c}0]^{(m)} = PA_i[\mathbf{a}0|\mathbf{c}0]^{(m)} + WP_i[\mathbf{a}0|\mathbf{c}0]^{(m+1)}$$
$$+ \frac{a_i}{2\xi}\left([\mathbf{a} - 1_i, 0|\mathbf{c}0]^{(m)} - \frac{\eta}{\zeta + \eta}[\mathbf{a} - 1_i, 0|\mathbf{c}0]^{(m+1)}\right)$$
$$+ \frac{c_i}{2(\zeta + \eta)}[\mathbf{a}0|\mathbf{c} - 1_i, 0]^{(m+1)}. \qquad (7.23)$$

The second relation is known as the Horizontal Recurrence Relation (HRR), also referred to as the Transfer Relation, and serves the purpose of "transferring" the angular momentum form centers **A** and **C** to centers **B** and **D**

respectively:

$$(\mathbf{a}, \mathbf{b} + 1_i | \mathbf{cd}) = (\mathbf{a} + 1_i, \mathbf{b} | \mathbf{cd}) + AB_i(\mathbf{ab}|\mathbf{cd}). \qquad (7.24)$$

The prefactors in (7.23) still depend on the exponents of the Gaussian functions, and therefore cannot be applied to ERIs over contracted functions. Prefactors in HRR depend on the geometric variables only, which are common for all primitive functions in a contraction. Hence, HRR *can* be applied to "contracted" ERIs. The computation strategy has to be adjusted accordingly:

1. For each combination of primitive Gaussians [**ab**|**cd**] do:
   (a) Compute $F_m(T)$;
   (b) Apply VRR to build primitive [**e**0|**f**0], where $\lambda_a \leq \lambda_e \leq \lambda_a + \lambda_b$, $\lambda_c \leq \lambda_f \leq \lambda_c + \lambda_d$;

2. Contract all [**e**0|**f**0] to form contracted (**e**0|**f**0) integrals;

3. Apply HRR to form contracted (**ab**|**cd**).

Head-Gordon and Pople's algorithm (HGP) outperforms the standard OS method for contracted integrals by virtue of shifting part of the workload outside the contraction loops. It is considered to be nearly optimal for uncontracted high angular momentum functions.

For contracted low-angular momentum classes there exist other recurrence relations which we will not consider here. Interested readers should refer to the following papers:
P. M. W. Gill and J. A. Pople, Int. J. Quantum Chem. **40** 753 (1991);
S. Ten-no, Chem. Phys. Lett. **211**, 3963 (1993);

# 8 Summary

Authors gratefully acknowledge Jason Gonzales for help with proofreading this document and useful comments. E.F.V. also thanks Tool for developing creative atmosphere in the room.

## 9   Appendix

Let us derive the explicit expression for the electron repulsion integral over primitive Gaussians using the Fourier transform for the $\frac{1}{r_{12}}$ factor:

$$\begin{aligned}
\phi_i(\mathbf{r}_1) &= x_{1A}^{l_a} y_{1A}^{m_a} z_{1A}^{n_a} \exp(-\alpha_1 r_{1A}^2) \\
\phi_j(\mathbf{r}_1) &= x_{1B}^{l_b} y_{1B}^{m_b} z_{1B}^{n_b} \exp(-\alpha_2 r_{1B}^2) \\
\phi_k(\mathbf{r}_2) &= x_{2C}^{l_c} y_{2C}^{m_c} z_{2C}^{n_c} \exp(-\alpha_3 r_{2C}^2) \\
\phi_l(\mathbf{r}_2) &= x_{2D}^{l_d} y_{2D}^{m_d} z_{2D}^{n_d} \exp(-\alpha_4 r_{2D}^2) \\
I &= \int \phi_i(\mathbf{r}_1)\phi_j(\mathbf{r}_1)\frac{1}{r_{12}}\phi_k(\mathbf{r}_2)\phi_l(\mathbf{r}_2) d\mathbf{r}_1 d\mathbf{r}_2 \\
\frac{1}{r_{12}} &= \frac{1}{2\pi^2} \iiint k^{-2} e^{i\mathbf{k}\cdot\mathbf{r}_{12}} d\mathbf{k} \quad\quad\quad (9.25)
\end{aligned}$$